\documentclass{sig-alternate}

\usepackage{color}
\usepackage{listings}
\definecolor{editorGray}{rgb}{1, 1, 1}
\definecolor{editorOcher}{rgb}{1, 0.0, 0.0} % #FF7F00 -> rgb(239, 169, 0)
\definecolor{editorGreen}{rgb}{1., 0.5, 0} % #007C00 -> rgb(0, 124, 0)
\usepackage{upquote}

\lstdefinelanguage{Csound}{
  morekeywords={instr, ins, chnget, outs, endin, min, max, step},
  morecomment=[s]{/*}{*/},
  morecomment=[l]//,
  morestring=[b]",
  morestring=[b]'
}
\lstdefinelanguage{JavaScript}{
  morekeywords={typeof, new, true, false, catch, function, return, null, catch, switch, var, if, in, while, do, else, case, break, instr, ins, chnget, outs, endin},
  morecomment=[s]{/*}{*/},
  morecomment=[l]//,
  morestring=[b]",
  morestring=[b]'
}

\lstdefinelanguage{HTML5}{
  language=html,
  sensitive=true,	
  alsoletter={<>-},	
  morecomment=[s]{<!-}{-->},
  tag=[s],
  otherkeywords={
  >,
	<!DOCTYPE,
  </html, <html, <head, <title, </title, <style, </style, <link, </head, <meta, />,
	</body, <body,
	</div, <div, </div>, 
	</p, <p, </p>,
	</script, <script,
  <canvas, /canvas>, <svg, <rect, <animateTransform, </rect>, </svg>, <video, <source, <iframe, </iframe>, </video>, <image, </image>, <input, </input
  },
  ndkeywords={
  charset=, src=, id =, width=, height=, type=, id= style=, type=, min=, max=, step=, rel=, href=,
  fill=, attributeName=, begin=, dur=, from=, to=, poster=, controls=, x=, y=, repeatCount=, xlink:href=,
  margin:, padding:, background-image:, border:, top:, left:, position:, width:, height:,
  transform:, -moz-transform:, -webkit-transform:,
  animation:, -webkit-animation:,
  transition:,  transition-duration:, transition-property:, transition-timing-function:,
  }
}

\usepackage[
  pass,% keep layout unchanged 
]{geometry}

\begin{document}

% Copyright
\setcopyright{waclicense}

%% DOI
%\doi{10.475/123_4}
%
%% ISBN
%\isbn{123-4567-24-567/08/06}
%
%%Conference
\conferenceinfo{Web Audio Conference WAC-2018,}{September 19--21, 2018, Berlin, Germany.}
\CopyrightYear{2018}

\title{WAAW Csound}
\numberofauthors{3} %  in this sample file, there are a *total*

\author{
% 1st. author
\alignauthor
       \name{Steven Yi}
       \affaddr{Rochester, USA}
       \email{stevenyi@gmail.com}
% 2nd. author
\alignauthor
       \name{Victor Lazzarini}
       \affaddr{Music Department\\ Maynooth University, Ireland}
       \email{victor.lazzarini@mu.ie}
% 3rd. author
\alignauthor
       \name{Ed Costello}
       \affaddr{Music Department\\ Maynooth University, Ireland}
       \email{edward.costello@mu.ie}
}

\date{\today}
\maketitle
\begin{sloppypar}
\begin{abstract}
This paper describes Web Assembly Audio Worklet (WAAW) Csound, one of the
implementations of WebAudio Csound. We begin by introducing the background to this current
implementation, stemming from the two first ports of Csound to the web
platform using Native Clients and asm.js. The technology of
Web Assembly is then introduced and discussed in its more relevant
aspects. The AudioWorklet interface of Web Audio API is explored,
together with its use in WAAW Csound. We complement this
discussion by considering the overarching question of support
for multiple platforms, which implement different versions of Web Audio. 
Some initial examples of the system are presented to illustrate 
various potential applications. Finally, we complement the paper by 
discussing current issues that are fundamental
for this project and others that rely on the development of a robust support for 
WASM-based audio computing. 
\end{abstract}
\section{Introduction}
%introduce Csound, Csound for Web Browsers,
%past work on Emscripten and PNaCl (ended)

The Csound sound and music computing system \cite{Csound2016book} runs
ubiquitously in a variety of platforms from supercomputers to embedded hardware.
Since 2013, it has also been ported to Web browsers \cite{CsoundWeb, lazzariniextending}, 
originally in two versions: one using an asm.js codebase and another as a 
portable Native Client (PNaCl) \cite{nacl,nacl2,pnacl} module  The first of these two implementations 
took advantage of two technologies, the Emscripten \cite{Zakai} compiler, and Web Audio. The second
version employed an ahead-of-time compiled bytecode module that was loaded by 
the browser as a plugin, and relied on the Pepper API as provided by the 
PNaCl SDK to implement audio input and output.

The pure JavaScript (JS) port of Csound could be deployed in any Web Audio-enabled
browser, whereas PNaCl Csound depended on the support for that technology,
which was limited to Chrome and Chromium-based browsers. However, it was
clear that the PNaCl approach was much superior from an audio programming
perspective. Not only it run very closely to native speeds, but also, since Pepper
provided a standard callback audio output interface, it was capable of much
better latencies (down to a 128-frame buffer size). Also, very importantly,
it ran audio processing on a separate thread, which was very robust and 
resilient to processing interruptions that might have caused drop-outs. As
far as audio programming on the Web was concerned, PNaCl still represents
a standard against which we can compare other solutions, including the one 
discussed in this paper.

With the Emscripten-compiled version, processing was of an order of magnitude
slower than PNaCl. Since it depended on Web Audio for output, it had to
run on its ScriptProcessorNode (SPN), which was neither designed for efficient 
processing, nor for low latency. Furthermore, the SPN has
a fundamental flaw in that it runs on the browser main thread, and therefore it
is prone to be interrupted by any graphics that is a little demanding. The only
advantage of this version was its wide support, which allowed code to run
on a variety of browsers and devices (including mobile). 

This was the situation until PNaCl was deprecated and its support discontinued,
following which the Csound project also stopped development of its PNaCl-based port and
frontend. More or less simultaneously, support for another compiler technology began
to be introduced. This was Web Assembly (WASM) \cite{Haas}, which, from the perspective
of Csound came as an alternative to its pure-JS port, using the same
SPN infrastructure, but possibly offering better computing
performance. 

More recently, after a long wait, the specification for the AudioWorklet (AW) was 
finalised in the WebAudio API, and support for it came on stream in the
Chrome browser. This offers an asynchronous alternative to the 
SPN, which is supposed to resolve the issues of 
sharing a thread with graphics/text display. This paper discusses 
Web Assembly-based Audio Worklet (WAAW) Csound:  an
implementation of the system for the browser platform, which 
takes advantage of these two technologies. 

The paper is organised as follows: we first detail
WASM and its use in the implementation of CsoundObj.js, the fundamental
API used to run Csound in a browser. We also discuss the porting of
csound.js, which is a frontend originally designed for PNaCl Csound,
which has been made to work with CsoundObj.js and WASM to provide
some compatibility with code written for PNaCl. We then move to introduce
AWs within the Csound context. We also tackle the issue of
multiple platforms, given the sketchy support currently enjoyed by this
component of the Web Audio API. The paper concludes with some
examples and a discussion of some up-to-date issues in 
Web Audio development vis-a-vis the technologies discussed here.

\section{Web Assembly}
%introduce Web Assembly, talk about CsoundObj
%and WASM Csound
WASM is a solution for the need of a low-level safe, fast, portable,
and compact code for the Web. It addresses the gap left by
the presence of JS as the sole natively supported
programming language in this platform. It has been developed
as a collaboration between the major browser vendors and
the online development community seeking a solution for
high-performance solutions. Among these, the case of
audio computing is listed as one of the prime targets for
WASM.

In the context of the Web, WASM is embedded in the
JS virtual machine. This takes the static representation
of a program, the WASM binary, and instantiates it
for execution. This may begin by invoking an
exported function from the binary instance.

WASM is both a bytecode format and a language with a
well-defined (and compact) abstract syntax  \cite{Haas}. This is
structured around a \emph{Module}, which is the WASM
binary containing definitions for \emph{functions},
\emph{globals}, \emph{tables}, and \emph{memories}.
WASM executes as a \emph{stack machine}, where
code for functions is a a sequence of instructions that
operate on an operand stack. 

Functions can be invoked through direct or indirect calls. In the latter
case, emulation of function pointers is possible, and
calls perform dynamic type checking. The indirect call
mechanism enables simpler support for dynamic linking.
Functions can also be exported or imported from/to a module, through
which a foreign function interface is implemented to
provide a communication with the external (embedding)
environment.

The WASM syntax includes the definition for four
basic types: two sizes of integer and IEEE 754
floating-point numbers each, with no distinction
between signed and unsigned integers. It allows for the
declaration of local and global variables of these
types.

Memory is provided by a linear array of bytes,
which is defined module-wise and can be grown
dynamically. WASM memory is defined to have
little-endian byte order. For security, it is 
completely disjoint from code space, as
well as the execution stack, so that exploits
or bugs have only a limited effect on an
running program.

Finally, an important feature of the WASM language 
is its representation of control flow, based on
the concept of \emph{structured control flow}.
Instead of allowing simple jumps, which may be unstructured and
irreducible, WASM provides structured blocks where branching
occurs as breaks, which can only take place
within a given scope given by the nesting of blocks. 
This has a number of benefits, including
single-pass validation and compilation.

In practical terms, WASM code can be 
generated by a C/C++ compiling environment,
for embedding in a JS virtual machine.
This bytecode is loaded, compiled, and invoked
via a JS API in three steps: 

% SYY - there's a Module.compile() or Module.compileStreaming(), 
% then a Module.instance() or Module.instantiateStreaming()
% Should we distinguish as four steps: load, compile, instantiate, use ? 
% VL - we could. I was basing this three steps from \cite{Haas}
% the reference paper on WASM.

\begin{enumerate}
\item A binary module is acquired from a source (disk, network);
\item Instantiated;
\item Exported functions can then be invoked.
\end{enumerate}

Implementations in JS engines take advantage
of a combination of ahead-of-time and just-in-time compilation
to translate bytecode efficiently. The API also provides means
of asynchronous loading and compilation of WASM modules,
through a promise mechanism.

Performance-wise, WASM has  been shown to perform
in many cases within 10\% of native code and in general,
within 50\% \cite{Haas}. It provides a significant improvement
over asm.js, in load time, performance, and executable
size. It is also on average 85.3\% of the size of a native x86\_64
binary. It appears to fill the necessary requirements for
high-performance applications such as realtime audio processing,
which would make it comparable to PNaCl (as far as
computation is concerned). Provided that support for
lower-latency audio IO is present in realtime-safe conditions, 
WASM appears to be an excellent solution for sound and
music computing on the web platform.

\subsection{WASM Csound}\label{wasmcsound}
% specific details about WASM Csound

WASM Csound is a port of the Csound library built with
the Emscripten cross-compilation tools. The core code 
has only one dependency, libsndfile, which is also built
for WASM and linked to Csound.  Together with the
core library, WASM Csound also includes the code for the CsoundObj,
which mediates access to the underlying Csound API. This is 
used in CsoundObj.js, where we interface with the Web Audio
API for audio IO, as discussed in \cite{lazzariniextending}. 

The WASM build is made up of two components, 

\begin{enumerate}
\item The WASM binary module (libcsound.wasm), which is
a static build of the Csound library, libsndfile, and includes
CsoundObj (as well as a couple of file listing utility functions).
\item Its JS interface (libcsound.js), generated at link time.
\end{enumerate}

The JS file contains the code that loads the
WASM code in the form of a Module object. Loading can
be synchronous or asynchronous and needs to be
determined at link time, when the JS interface
is generated. Depending on how we intend to use the code, 
we might need to chose one or the other loading method.

At the moment, the Csound API is exclusively accessed
through the exported functions in CsoundObj.c, which
wraps the relevant parts of the API. In the future, it
might be a consideration to provide access to the underlying
API, but this has to be considered more carefully. The 
CsoundObj.js file provides the access to the exported
functions and connects to Web Audio for
audio IO, as mentioned above. In the cases where AW is not
available, this uses SPN as we had
done before with asm.js \cite{lazzariniextending} (see Sec. \ref{multiple}
for details).

\subsection{csound.js}
% talk about csound.js as a port of the PNaCl
% frontend to WASM Csound

With the demise of PNaCl, Csound support for that 
technology has also been retired. In order to maintain 
compatibility with JS code that used PNaCl Csound,
a port of its frontend, csound.js, has been developed. 
Built on top of the CsoundObj API, it provides means
to run a single Csound engine per web page, send
Csound code for compilation, access to its software bus
channels, and MIDI input. It also includes support for accessing
the sandboxed filesystem used by Csound, for the loading
of source, audio, MIDI, and other binary files.
It provides an alternative to the direct use of CsoundObj, 
simplifying the code for a number of use cases. Depending on their
requirements, user applications 
can be written using the functionality provided by csound.js or by 
CsoundObj.js

\section{Audio Worklets}\label{aw}
% introduce audio worklets and discuss other
% experiences (cite Jari, Oli, Stephane)
% explain chosen approach and design

Audio Worklet (AW) is a new interface provided by the Web Audio API \cite{Webaudio}
for the developers to provide scripts establishing custom audio nodes,
which will be executed by a worker thread, rather than the main thread
(as in the case of ScriptProcessor nodes).  The potential for this new
interface, if implemented correctly, is to provide a more robust realtime audio 
support for systems such as Csound, which employ Web Audio mostly
for audio IO, or as an extension of the built-in unit generators provided
by that API. 

\begin{figure}[htp]
\begin{center}
\includegraphics[width=0.9\columnwidth]{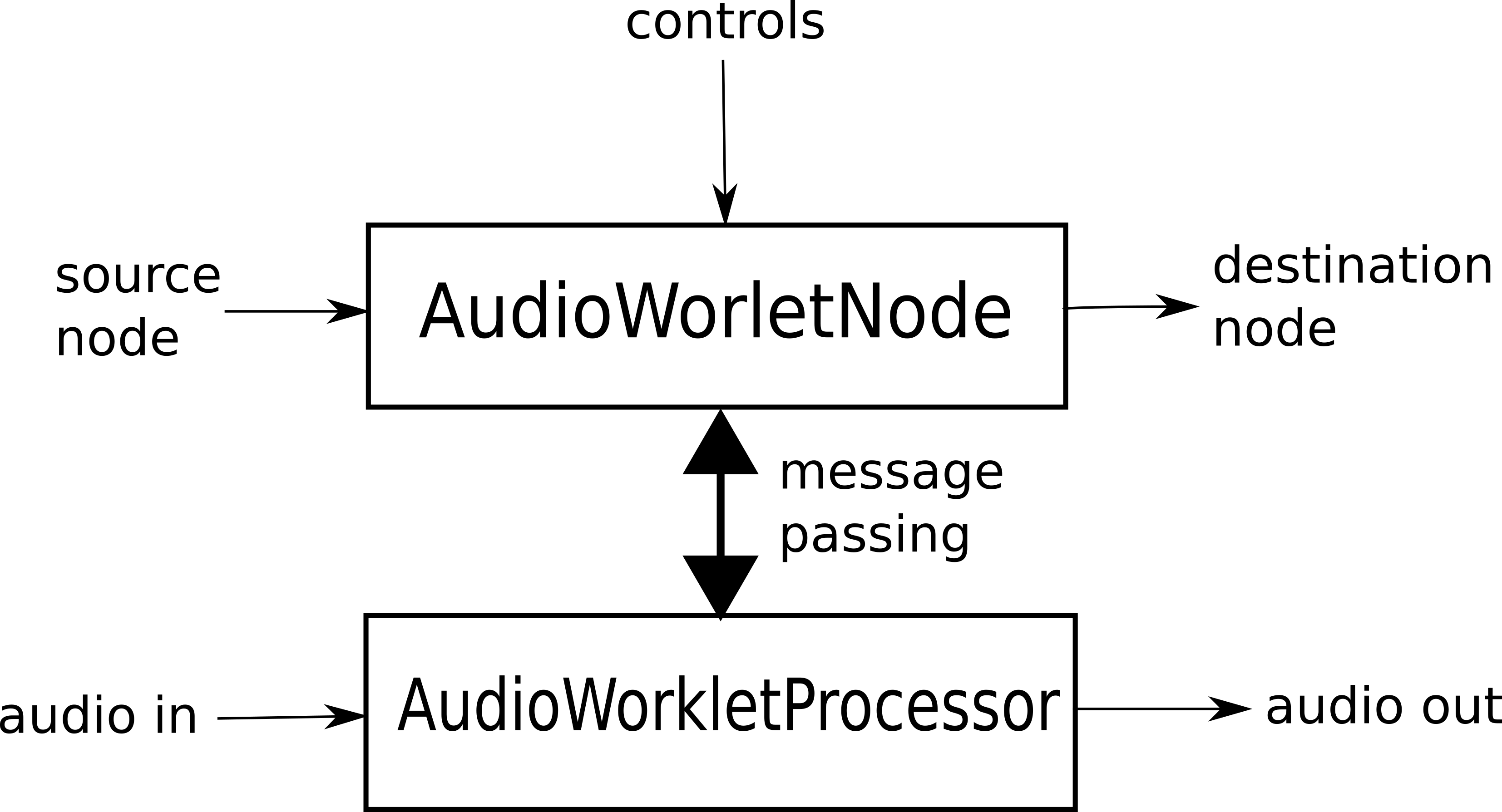}
\caption{AudioWorkletNode and AudioWorkletProcessor}
\label{fig:awpawn}
\end{center}
\end{figure}

For an AW to be set up, a pair of objects  must be defined, namely an 
AudioWorkletNode (AWN) and an AudioWorkletProcessor (AWP).  The latter contains the implementation 
of the audio processing code, and the former provides
an interface for the main global scope, as a custom AudioNode. Communication
with the global scope is implemented via asynchronous message passing,
and audio processing is synchronous with the rest of the audio graph
defined in an application (Fig. \ref{fig:awpawn}).

\subsection{CsoundObj.js integration}\label{integration}

In order to integrate this new AW functionality,
the CsoundObj API has been redesigned to have a 1:1 relationship with a WASM-backed
Node implementation, either AWN or SPN, as detailed in Sec. \ref{multiple}. 
The choice of backend defaults to the best performing option at runtime (AW if
available, otherwise fall back to SPN) though may be overridden if
a specific implementation is required. 

Developers may work with CsoundObj without understanding the WebAudio API to create 
music applications. In this case, CsoundObj will manage the audio graph of nodes for 
the developer. Using CsoundObj in this manner follows usage of other CsoundObj
API implementations found in Java on Android and in Objective-C and Swift on iOS.

Developers may also access the Csound Node (either AudioNode or SPN)
that is backing the current CsoundObj instance. In this scenario, developers 
use CsoundObj simply as a Node factory, and take the responsibility for
assembling and maintaining the audio graph themselves. 

These two approaches to CsoundObj API usage allow for covering a wide range of
use cases.  We hope this encourages Csound users to explore making web applications
with their Csound knowledge, and Web Audio API users to explore extending their systems
with Csound-backed Nodes. 

\subsection{Setting up and registering the AWP}

The \lstinline{addModule()} method of the \lstinline{audioWorklet} object in the AudioContext is used to set up the AWP in three primary steps into the 
\lstinline{AudioWorkletGlobalScope}: 

\begin{enumerate}
 \item Load the WASM binary data
\item Load the libcsound.js WASM module loading script. This is generated using Emscripten (see Sec. \ref{wasmcsound}) with the MODULARIZE=1 option.
\item Load the CsoundProcessor.js. This is the JS code that loads the WASM module from the WASM binary, reads native functions to call from JS using cwrap(), then defines and registers the Csound AWP. 
\end{enumerate}

For WASM-backed AWs, there has been difficulty in trying to find ways to get the WASM binary data to the AWP thread. Following the work done for WebAudioModules and Faust and in contact with their authors, we have explored two primary ways to achieve this: encode binary data as JS code that instantiates a \lstinline{Uint8Array} that is passed directly to the module loader; encode binary data as base64-encoded string embedded in JS code that is decoded to binary at runtime before being passed to the module loader. 

Table~\ref{fig:encoded_wasm} shows a comparison of various properties between the two encoding methods.\footnote{The generated WASM was compiled using Emscripten SDK 1.37.37. The speeds were reported using Chrome 66 on Windows 10, Surface Pro 4, Core i7-6650U CPU @ 2.20 GHz.} While on-disk size was significantly larger for the Uint8Array encoding, the compressed size delivered over network was only 100 kB difference.  The load time\footnote{Measured using 
new \lstinline{Date().getTime()} at start and end of JS files, as this method was available both on the main thread and AWP thread.} 
shows an order of magnitude difference due to the runtime decoding of base64 back to binary.  Factoring in that the cost for decoding base64 was required every load, but downloading the JS was likely to be a one time cost due to caching, we settled on using the Uint8Array implementation (see Sec. \ref{wasm-issues} for further details). 

\begin{table}
\caption{Comparison of base64 and Uint8Array encoded WASM Properties}
\label{fig:encoded_wasm}
\begin{center}
\begin{tabular}{ |c|c|c|c| } 
  \hline
  Property & base64 & Uint8Array \\
  \hline
  File Size & 3.9 mb & 8.3 mb \\ 
  Network Size & 1.5 mb & 1.6 mb \\ 
  Load Time & 73 ms & 6 ms \\ 
  \hline
\end{tabular}
\end{center}
\end{table}

% SYY - Stopping here for moment to retest using Uint8array version to see how that performs and what to write up here

\subsection{Instantiating the AWN/AWP}

When a CsoundNode is created on the main thread, the corresponding AWP (CsoundProcessor) is also created. When the AWP\'s constructor is executed, a new native Csound instance is created and ready for use.  At this point, the AWP is ready for processing and the AWN is ready for connection in an audio graph.  If the native Csound instance is not in a running state, the AWP \lstinline{process()} method is a non-op.  When Csound is in a running state, the \lstinline{process()} method transfers AWP input to Csound, executes Csound by calling the \lstinline{CsoundPerformKsmps()} API function, then transfers audio output from Csound to the output buffer channels (Listing \ref{process}).

\begin{lstlisting}[language=javascript, caption={CsoundProcessor \lstinline{process()} method}, label={process}]

process(inputs, outputs, parameters) {
 if(this.csoundOutputBuffer == null ||
     this.running == false) { return true;}
 let input = inputs[0];
 let output = outputs[0];
 let bufferLen = output[0].length;
 let csOut = this.csoundOutputBuffer;
 let csIn = this.csoundInputBuffer;
 let ksmps = this.ksmps;
 let zerodBFS = this.zerodBFS;
 let cnt = this.cnt;
 let nchnls = this.nchnls;
 let status = this.status;
 for(let i = 0; i < bufferLen; i++, cnt++) {
  if(cnt == ksmps && status == 0) {
    status = Csound.performKsmps(this.csObj);
    cnt = 0;
  }
  for(let channel = 0; 
        channel < input.length; channel++) {
    let inputChannel = input[channel];
    csIn[cnt*nchnls + channel] = 
       inputChannel[i] * zerodBFS
  }
  for(let channel = 0;  
        channel < output.length; channel++) {
    let outputChannel = output[channel];
   if(status == 0)
     outputChannel[i] = 
      csOut[cnt*nchnls + channel]/zerodBFS;
   else
     outputChannel[i] = 0;
   } 
 }
 this.cnt = cnt;
 this.status = status;
 return true;
}
\end{lstlisting}

% SYY - In the above, may be simpler to just say AWP delegates to Csound when Csound is in running state or does a no-op

\subsection{Using Csound through the AWN/AWP}

As explained in Sec. \ref{integration}, developers may employ the CsoundObj interface to work with Csound. This involves operations such as setting 
Csound engine options, compiling Csound code, starting the Csound engine, and sending control data and events to a running Csound instance. 
Methods for these operations delegate to the AWN, which in turn communicates asynchronously with the AWP, where they are executed.
Alternatively, a CsoundNode may be created with the static method \lstinline{CsoundObj.createNode()} and used as another audio node in the system.

%\subsection{CsoundNode}

\section{Multiple Platforms}\label{multiple}

Prior to  the introduction of AW, WebAudio Csound was released in two implementations. The original one packaged the CsoundObj API that used SPN and a native C libcsound compiled into asm.js. The second form, WASM Csound, as discussed in Sec \ref{wasmcsound}, used an identical API but was backed by a WASM libcsound module. Developers could download one or the other release depending upon the technology they wanted to support. Developers could also write code to, at runtime, choose which implementation to load. However, it was the responsibility of the developer to write their own loading mechanism if they wanted to have runtime implementation switching. 

With the arrival of AW and the ubiquitous availability of WASM, the WebAudio Csound release has dropped the asm.js form and now includes the two supported forms: the WASM-backed SPN (aka WASP Csound) or WASM-backed AW (aka WAAW Csound).  The CsoundObj API also now includes all code to detect the platform and capabilities at runtime and choose to which implementation to use.

We have tested web applications using the current WebAudio Csound on various desktop and mobile platforms. The AW version is properly loaded in applications served over HTTPS to browsers that support AW (tested with Chrome Stable 66 and Chrome Canary 68 on Windows 10 and MacOS; Chrome Beta 66 on Android). The SPN version is properly loaded when the web application is served over HTTP or when the browser does not support AW (tested Firefox on Windows, Linux, MacOS, and Android; MS Edge on Windows 10)\footnote{WebAudio Csound does not properly run on iOS at this time though the authors have found many other WebAudio applications do not currently function on this platform. We consider it an ongoing problem to solve.}.

% discuss how to approach the sketchy
% implementation of Audio Worklet
% across platforms
% and how we provide support
% for both SPN and
% CsoundNode in CsoundObj.

\section{Examples}
% show a simple example 
% to demonstrate use
A set of examples demonstrating WAAW Csound have been provided as
straight ports of original WASM and PNaCl Csound pages, which use csound.js and a plain HTML 5
interface. These demonstrate basic interaction and synthesis, range from simple sound 
tests, to interactive synthesis examples, as well as MIDI and CSD file players. 
They can be found at

\begin{center}
\url{https://waaw.csound.com} 
\end{center}

As discussed in Sec. \ref{multiple}, these examples will use AW if available (WAAW), falling back to SPN (WASP) if not, hence
they can be used for some head-to-head comparisons between the two implementations.

\subsection{Glowing Orbs}

``Glowing Orbs''\footnote{\url{https://kunstmusik.github.io/csound-p5js}} is an example web application project that demonstrates building a generative multimedia program using Csound for audio and p5.js\footnote{http://p5js.org} for animation. It is an example program designed to function as a template for users to download, customise, and use to create their own interactive, multimedia web programs. 

Originally developed with WASM Csound using SPN, the program ran fairly well but suffered from two primary issues due to both audio and animation (via p5.js) being rendered in the main thread. Firstly, the audio could break up if the animation took up too much time. This was evident during development and facilitated changes to reduce the number of orbs to use on the screen to ensure there was enough rendering time for the audio to avoid breakups. Secondly, the video frame rate could slow down if the audio took up too much time.  This was clearly seen on more resource-constrained platforms, such as Android.  

Switching to WAAW Csound, we found that using the AW path successfully provided the expected benefits to render audio separately from the main thread. Audio rendered well and video animation was much smoother compared to the original implementation.  We have observed a similar improvement in other examples
that used moderate amounts of graphics rendering, such as for instance in the display of console messages from Csound.

\subsection{live.csound.com}
% introduce and 
% discuss live.csound.com

The live.csound.com\footnote{\url{https://live.csound.com}} website is a Progressive Web Application (PWA)\footnote{\url{https://developers.google.com/web/progressive-web-apps/}} that provides a full Csound livecoding environment in the browser. Users can program in Csound code using the included livecode.orc framework\footnote{\url{https://github.io/kunstmusik/csound-live-code}} as well as employing any other standard Csound programming practices. The web application may be installed to users' devices, such as Android phones, and run offline.  

Using the application with WAAW Csound, we found that we no longer had to worry about interactions with the text editor interfering with sound processing.  This matched our expectations for the benefits of AW design over SPN as well as matched the behaviour we enjoyed previously with PNaCl Csound. We also verified that the AW implementation worked equally well when running offline. Our initial experiences with integrating WAAW Csound with live.csound.com have been positive and we look forward to exploring the area of PWA music applications further using Csound.

\section{Current Issues with AW}
% discuss the difficulties with
% async vs. sync loading 
% plus support in different
% browsers and iOS problems

 As noted in Sec. \ref{aw}, AW is a very recent addition to the Web Audio API. As such, at the time of
writing, they have only been implemented by the Chromium project and
deployed for the first time in Chrome 66. This first implementation has a
number of teething issues, which have been the subject of discussion between
a group of early adopters of the technology (which includes these authors) and
Chromium project developers. The major issues can be classed into two
categories: (a) audio computing, and (b) WASM integration.

\subsection{Audio Computing}\label{audio-issues}

Early adopters of AW have noticed intermittent drop-outs in the audio
stream. Since the promise of the technology was exactly to provide
a robust and resilient environment for JS (and now WASM)-generated
audio, this is clearly a significant issue. Two related difficulties seem to
have been identified:

\begin{enumerate}
\item Thread priority.
\item Garbage collection.
\end{enumerate}

In the V8 engine provided by the Chromium project, there are four levels
of priority: BACKGROUND, NORMAL, DISPLAY, and REALTIME\_AUDIO.
The thread in which an AW runs is scheduled with DISPLAY priority, whereas
native nodes run at REALTIME\_AUDIO priority. This has been identified
as less than ideal, and a proposal to move AW to REALTIME\_AUDIO has
been put forward. There is resistance to this, arising from security concerns,
so it is not clear at the time of writing whether the proposal will be accepted.

Running at DISPLAY priority might still be a reasonable proposition if
it were not for the fact that the JS engine garbage collector 
(called Oilpan in V8) seems to be getting in the way of audio computing.
As it stands, the AW thread appears to invoke garbage collection 
on a regular basis; moreover, this needs to be synchronised with 
garbage collection in other threads. A programmatic example 
that places garbage collection pressure on the main thread,
prepared by these authors, has demonstrated clear break-ups
in the audio stream when computing a simple sine wave in
WAAW Csound.

It is reported that audio buffers are allocated at every audio
cycle \cite{HongChan1}. Allocation is from a memory pool, so in itself might
not be too problematic  but it means that
garbage collection is involved in the process. From the point
of view of realtime time audio  \cite{Bencina}, this is really unsatisfactory,
but from the perspective of implementors, nothing in the
specification explicitly prohibits it. 

The major question to be asked is whether the Web Audio design 
for AW should have included more stringent specifications for 
realtime operation. In fact, we should have expected a
level of realtime safety that is comparable to native node
implementation. We would for instance like to have a lock-free
AW thread, which given the level of difficulties involved
seems at the moment to be unachievable.

In the end, it is accepted that the Web platform, as
implemented by a JS engine, is not designed
as a realtime system \cite{HongChan2}. All we can
do is mitigate the issues that arise from a general-purpose
computing environment where audio is not anywhere
the main concern.

The regrettable aspect of this situation is that five
years ago we already had an audio computing 
environment for browsers that offered realtime
safe operation, in the form of PNaCl modules using
the Pepper audio API. At the first Web
Audio Conference in 2014, these and other authors demonstrated
systems that took advantage of that technology to
provide robust audio computing solutions. At that
time, AW (more specifically, its design predecessor AudioWorkerNode) 
was a near-future promise that would prospectively solve many of the existing problems
for non-builtin audio processing in Web Audio. More than 
four years later, this goal has not yet been fulfilled,
and an alternative has been discontinued. Such issues
with the Web Audio API development have been already
noted elsewhere \cite{LazzariniYi}. However, it remains the only 
means of audio IO provided in the web platform for foreseeable future,
therefore we are committed to contributing to the search for workable solutions.

\subsection{WASM support}\label{wasm-issues}
% please discuss what is needed to improve support
% for WASM, e.g. loading code etc 

Using WASM with AW is straightforward, once the binary module has
been instantiated and available to use within the AWP thread. However, loading
and compiling the WASM binary is more complicated than we would have liked. 

Ideally, we would deploy a single WASM binary and loader script that could be
used for both AW and SPN versions of Csound. The idea is that we might initiate
loading, compilation, and instantiation of the WASM module from the binary
either in the main thread for SPN or AWP thread for AW.  

However, this is not currently possible as we found two conflicting
requirements.  Firstly, compiling WASM within the AWP thread requires that
Module compilation be done synchronously. Secondly, compiling WASM for use with
SPN requires it be done asynchronously (e.g., Firefox 59.0.2 on Android throws
an error if synchronous WASM compilation is attempted from the main thread).
As a result, until all of the major browsers support AW, we will need to
package two WASM binaries and JS loader scripts to make our WebAudio Csound-based
applications operate as widely as possible.

Next, because code within the AWP thread cannot use standard JS loading methods 
like \lstinline{fetch()}\footnote{See \url{https://github.com/WebAudio/web-audio-api/issues/1439}
for more information about the reasons for this.}, we had to consider how to get the
binary data to the AWP thread for it to compile and instantiate the WASM Module object.
Following the work of the community, we went with a method to encode the WASM binary
into Javascript code that creates a \lstinline{Uint8Array}.  The generated JS file is then used
with the \lstinline{addModule()} method of the \lstinline{audioWorklet} object to have that loaded in the AWP 
thread. The resulting variable is then used with the WASM loader script for libcsound
that results in synchronously compiling and instantiating the Module from the binary data.

Other loading methods have been proposed. One method is to first instantiate an AWN/AWP and
then transfer WASM binary data or compiled module to the node through its MessagePort.  This 
might work but complicates the loading/instantiation as one has to split construction of the
AWN/AWP into two phases: one phase when the constructor is called and another phase after 
the node has received WASM data. This is not a solution we found attractive.  

Another solution is to pass WASM data to the AWP constructor through the AWN
\lstinline{processorData} constructor argument. This is reported to be
working\footnote{\url{https://bugs.chromium.org/p/chromium/issues/detail?id=808848}}
in Chrome Canary (an unreleased, development version of Chrome) but requires a
special flag to allow it to work. At this time, we are monitoring the status of
this change and its presence in upcoming release versions of Chrome.  

For the short term, we have settled on using the \lstinline{Uint8Array} JS solution. This
solution works well in desktop Chrome 66 browsers, Android Chrome 66, and in
Chrome OS 66. The process to encode the WASM into JS is, in our opinion, a
hack to deal with an awkward situation, but we will continue to use it until a
better solution becomes commonly available. 

\section{Conclusions}
% summary, future prospects

This paper described the implementation of Csound under WASM
and AW. Building on five years of previous work,
we have been able to provide a generally stable and usable implementation,
which we will hope will be improved as some of the key hurdles in AW
implementation get overcome. Despite the difficulties encountered with
the early implementation of the interface in the Chrome project, we
found that there is good potential for development. Taking PNaCl+Pepper
as a standard for audio programming support on the Web Platform,
we have found that AW is going in the right direction, providing a
much improved environment when compared to the original SPN.
In particular, we have found that the combination of audio and
graphics has had significant gains in performance. 

Finally, we would like to note that Csound is free software, and the source code, plus all building scripts, 
is available publicly at

\begin{center}
\url{https://csound.com} 
\end{center}

The WebAudio Csound implementation as described in this paper has
been first introduced in version 6.11. 

\section{Acknowledgments}

The authors would like to thank Jari Kleimola, Oli Larkin, and St\'{e}phane Letz, early
adopters of the WASM+AW technology, for the useful discussions and sharing of
experiences. We would also like to note the support of Hong Chan from
the Chromium project, and his patience in listening to our requests
and suggestions. Finally, we would also like to thank Henri Manson
who contributed to the WASM Csound port effort.

\bibliographystyle{abbrv}
\bibliography{sigproc}  % sigproc.bib is the name of the Bibliography in this case

%\balancecolumns % GM June 2007

\end{sloppypar}
\end{document}